\documentstyle[pre,aps,twocolumn,epsfig]{revtex}
\begin{document}
\draft

\title{Simulated coevolution in a mutating ecology}
\author{J.S. S\'a Martins}
\address{Colorado Center for Chaos and Complexity, CIRES, 
University of Colorado, Boulder, Colorado 80309}

\maketitle

\begin{abstract}
The bit-string Penna model is used to simulate the competition between an
asexual parthenogenetic and a sexual population sharing the same
environment. A new born of either population can mutate and become a part
of the other with some probability. In a stable environment the sexual
population soon dies out. When an infestation by fastly mutating
genetically coupled parasites is introduced however, sexual reproduction
prevails, as predicted by the so-called Red Queen hypothesis for the
evolution of sex.

\end{abstract}

The question of why sexual reproduction prevails among an overwhelming
majority of species has resisted a century-long investigation. It is clear
that its appearance is rather ancestral for metazoan, multicellular
animals. Recombination probably originated some three thousand million
years ago, and eukaryotic sex one thousand million years ago. But the
mechanisms through which simple haploid organisms mutated into diploid
sexual forms (the origins of meiosis and the haploid cycle) remain one
of the great puzzles of evolution theory \cite{jms}. We can speculate
about these origins, but cannot test our speculations. In contrast,
selection must be acting today to maintain sex and recombination. We have
to concentrate on maintenance rather than origins, because only thus can
we have any hope of testing our ideas.

From a theoretical point of view, the selective advantages of a sexual
population over a simple asexual one are well understood, and could be
established through a variety of approaches. The bases for these
advantages derive first from the covering up, by complementation, of 
deleterious genes, then from the ability to recombine genetic material, 
which haploid asexual reproduction lacks. But serious difficulties arise 
when sexual reproduction is compared with meiotic parthenogenesis, which is 
a kind of diploid asexual regime that also involves genetic recombination. 
In this case, the two-fold advantage of not having to produce males could 
give parthenogenetic populations the upper hand against a competing sexual 
variety, since recombination is present in both. 

The theoretical problem thus posed to evolutionary biologists is indeed
very difficult. The observation of this competition in natural habitats is
not feasible, in general. One must rely on very indirect and often
questionable data. It should be pointed out that this is a feature shared
by many of the most important problems in the theory of evolution. This 
situation is certainly one of the main reasons for the recent
boost of activities in physics directed towards biology, since the same
methods and approaches could prove once again fruitful \cite{gabriel}. In
particular, physicists have pioneered in the use of techniques derived
from the availability of powerful low-cost
 computers to fulfill the lack of and complement experimentation
\cite{landau}. Computer simulations of natural systems can provide much
insight into their fundamental mechanisms, and can be used to put to a
test theoretical ideas that could be otherwise viewed as too vague to
deserve the status of scientific knowledge. The scientific literature of
this decade is strong testimony to the success of this approach, in
various and apparently disconnected fields; for a recent, although
partial, survey of nonphysical applications I direct the reader to Ref.
\cite{evolution}. It is against this background that this work is being
reported. 

Investigations of evolutionary problems by physicists have in fact boomed
in the last few years. In what concerns biological aging, this boom can be 
traced back to the introduction of Penna's bit-string model \cite{tjpp} 
which was quickly adopted as referential in most studies, as a sort of 
``Ising model'' of the field. Its simplicity and early successes in 
reproducing observed features of real populations, such as the Gompertz law 
\cite{bjp}, the Azbel phenomenology \cite{azbel} and the catastrophic 
senescence of semelparous animals \cite{rc}, unleashed a burst of efforts in 
its application to a number of different phenomena. In some simple cases, 
analytical solutions could even be provided, shedding some light on
the simulation results; recent reviews can be found in Ref. \cite{suzana}, 
and an extensive list of references in Ref. \cite{evolution}. 

Other models have been examined in the recent literature. Of particular
interest is the Redfield model \cite{redfield}, which assumes a constant
population and does not allow mutational meltdown. 

The Penna model has also been used to address the problem in question here, 
namely, the reasons for the maintenance of sexual reproduction
\cite{why}. Simulations with this model showed that a genetical
catastrophe could eliminate a parthenogenetical population, whereas
sexually reproducing species survived. This result had the merit of
pointing out a measurable effect of the greater genetic diversity created
by sex, but the occurrence in nature of catastrophes of that kind seems
rather unlikely. Its introduction in the simulation is prone to be
thought of as too artificial to be convincing. Another drawback should
also be mentioned. All previous comparisons between sexual and asexual
reproduction have come from results where each population evolved by itself, 
as if each one lived in a separate environment. This is
biologically improbable, and a simultaneous study of both regimes, sharing
the same resources, is needed. This has been already done, in the context
of the Redfield model. There, it was shown that sexual reproduction has
a short-term advantage over the haploid asexual regime if the female
mutation rate is high enough \cite{stauffer}. 

The present work represents a step in the direction of endowing the
artificial world in which the populations evolve in Penna model
simulations with more realistic features. Its purpose is again to verify
measurable effects of sexual genetic diversity when compared to
parthenogenetic reproduction. The simulations here reported are based on 
the recent field work of biologist Lively, in which he observed the effect
of parasitic infestation of a freshwater snail's ({\it Potamopyrgus
antipodarum}) natural habitat on its dominant reproductive regime
\cite{lively}. This observation can be considered as an illustration of
one of the theories in the debate over the reasons for the prevalence of
sex, the so-called ``Red Queen'' hypothesis. This is a variation of the idea
that sex serves to assemble beneficial mutation, and so creates a
well-adapted lineage in the face of a rapidly changing environment.
Because parasites adapt to the most common host genotype, evolution will
favor hosts with a rare combination of resistant genes. Thus, the Red Queen
predicts that selection will favor the ability to generate diversity and
rare genotypes, exactly the abilities conferred by sex and recombination. 

I describe in what follows the model used in the simulations, the Penna
bit-string model with recombination, and the representation of a parasitic
infestation in its context. 
The genome of each (diploid) organism is represented by two computer
words. In each word, a bit set to 1 at a position (``{\it locus}'')
corresponds to a deleterious mutation; a ``perfect'' strand would be
composed solely of zeros. The effect of this mutation may be felt 
at all ages equal to or above the numerical order of that {\it
locus} in the word. As an example, a bit set to one at the second position
of one of the bit-strings means that a harmful effect may become present
in the life history of the organism to which it corresponds after it has
lived for two periods (``years''). The diploid character of the genome is
related to the effectiveness of the mutations. A mutation in a position of
one of the strands is felt as harmful either because of homozygose or
because of dominance. For the former, a mutation must be present in both
strings at the same position to be effective. The concept of dominance on
the other hand relates to {\it loci} in the genome in which a mutation in
just one strand is enough to make it affect the organism's life. The life
span of an individual is controlled by the amount of effective mutations
active at any instant in time. This number must be smaller than a
specified threshold to keep the individual alive; it dies as soon as this
limit is reached. 

Reproduction is modeled by the introduction of new genomes in the
population. Each female becomes reproductive after having reached a
minimum age, after which it generates a fixed number of offspring at the
completion of each period of life. The meiotic cycle is represented by
the generation of a single-stranded cell out of the diploid genome. To do
so, each string of the parent genome is cut at a randomly selected
position, the same for both strings, and the left part of one is combined
with the right part of the other, thus generating two new combinations of
the original genes. The selection of one of these complete the formation
of the haploid gamete coming from the mother. 

The difference between sexual and parthenogenetic reproduction appears at
this stage. For the first, a male is selected in the population and
undergoes the same meiotic cycle, generating a second haploid gamete out
of his genome. The two gametes, one from each parent, are now combined to
form the genome of the offspring. Each of its strands was formed out of a
different set of genes. 

For parthenogenesys, all genetic information of the offspring comes from a
single parent. Its gamete is cloned, composing an homozygous genome for
the offspring. For both regimes, the next stage of the reproduction
process is the introduction of $m$ independent mutations in the newly
generated genetic strands. In this kind of model it is normal to consider
only the possibility of harmful mutations, because of their overwhelming
majority in nature. For sexual populations, the gender of the newborn is
then randomly selected, with equal probability for each sex. 

A last ingredient of the model is a logistic factor, called the Verhulst
factor, which accounts for the maximum carrying capacity of the
environment for this particular (group of) species. It introduces a
mean-field probability of death for an individual,
 coming from nongenetic causes, and for computer simulations has the
benefit of limiting the size of populations to be dealt with. 

The passage of time is represented by the reading of a new {\it locus} in
the genome of each individual in the population(s), and the increase of
its age by 1. After having accounted for the selection pressure of a
limiting number of effective (harmful) 
 mutations and the random action of the Verhulst dagger, females that have
reached the minimum age for reproduction generate a number of offspring.
The simulation runs for a prespecified number of time steps, at the end
of which averages are taken over the population(s).
A more detailed description of the standard Penna model can be found in
Ref. \cite{evolution}, together with a sample computer code that
implements its logic. 

The extension of the original Penna model to simulate the coevolution of
populations is rather straightforward. In this Rapid Communication, I focus 
on the
coevolution of different varieties of the same species, sharing the same
ecological range. This implies that the maximum carrying capacity relates
to the total population, summing up all varieties. I use this simple
extension to study the effect of introducing a small probability $p$ for a
mutation to transform offspring of one variety into the other. This
implies an extra stage for the reproduction logic. After a new-born
genome is generated from the sexual population, it mutates into a
parthenogenetic female and become part of the asexual population with
probability $p$. Accordingly, the offspring of an asexual female can
mutate to the sexual form, with the same probability $p$; if it does, a
gender is randomly chosen for it. 

Further extension is needed to simulate the conditions of a parasitic
infestation. I chose a very simple strategy to reproduce Lively's
observation. Parasites are represented by a memory genetic bank of a fixed
size $M$. At each time step, every female of the species establish contact 
with a fixed number $P$ of elements of the parasite population. If the 
female's genome is completely matched by the genome already memorized by the 
parasite, she loses its ability to procreate. This simulates the action of 
the parasite {\it Microphallus} on the fresh water snail: this trematode 
renders the snail sterile by eating its gonads. The parasite memory bank has, 
on the other hand, a dynamic evolution. If an element of the parasite 
population contacts the same genome a certain number of times $n$ in a row, 
this particular genetic configuration is ``learned'' by the parasite, and 
turns it active against this genome, until a new pattern is found 
repetitively. Note that the same genome can be present in a number of 
different females, so the action of the parasite is not restrained by being 
randomly ``chosen'' by the same female the required number of times. Rather, 
the effectiveness of this simulated parasitic infestation is an indirect 
measure of the genetic variability (actually, of the lack of this 
variability) within the female population of each variety.

It must be remarked that this is not the only possible choice for the 
dynamic evolution of the parasite population. Nonetheless, it captures the 
essential features of parasitism; in particular, parasites that match 
frequently occurring female's genomes will spread out through the population, 
showing that the above dynamical rules succeed in giving them an effective 
handicap.

All the species, the two varieties of the snail and the parasites, 
evolve as fast as they can, but neither gets far ahead, hence the theory's
name, after the Red Queen's remark to Alice in Wonderland: ``It takes all
the running you can do, to keep in the same place.'' The extra constraint
that the two varieties of snail are competing for the same resources of
the environment may eventually cause one of them to collapse. And Lively's
observations match this expectation: environments with few parasites tend
to have mostly asexual snail populations, while a larger number of
parasites is correlated to a predominant sexual variety. This pattern
strongly suggests that the parasites prevent the elimination of sexual
populations. 

\begin{figure}[htb]
\centerline{\psfig{file=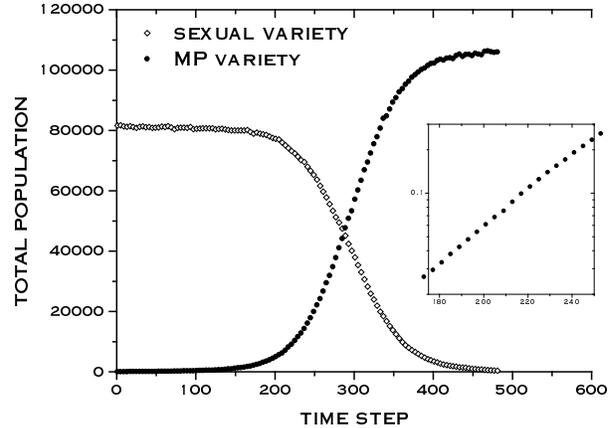, width=9cm, angle=0}}
\caption{Total population of each variety is plotted against time, 
measured in number of Monte Carlo steps.
After a short transient, the parthenogenetic (MP) variety dominates. The
inset shows a semilog plot of the fraction of asexual females in the
population as a function of time, exhibiting the expected exponential
increase. See text.}
\label{2fold}
\end{figure}

The two-fold short-term disadvantage of sexual reproduction is the object
of the first result shown. The initial population is composed of a single
sexual variety in which each reproductive female gives birth to a
parthenogenetic female with a small probability $p = 0.0001$. The minimum 
age for reproduction $R = 10$ and the birth rate per female $b = 1$ are 
fixed, and do not suffer the effects of mutations. Figure 
\ref{2fold} shows the time evolution of the total population of each
variety. In this run, the environment is kept fixed; there is no parasite
infestation. The burden of having to produce males in the population is
responsible for the quick establishment of a dominant asexual variety. The
inset shows the time evolution of the fraction of asexual females in the
total population once a parthenogenetic lineage appears for the first
time. A simple argument that underlies this result goes as follows
\cite{jms}: suppose there are $A$ asexual females and $S$ sexual ones 
(and also $S$ males) that have already reached reproduction age in a
generation. In the next one, there will be $bA$ and $bS/2$ of each variety; 
the factor of $2$ of the latter comes up because of equal probabilities
to give birth to males and females. The fraction of parthenogenetic
females in the total population increases from {\large $\frac{A}{A + 2S}$}
to {\large $\frac{A}{A + S}$}; when $A$ is small, this is a doubling in each
generation. This reasoning presupposes an unbounded increase in the total
population and no overlapping between generations, which are not realistic
assumptions in general. Instead, one should expect a smaller factor for
this growth. The simulation shows clearly this exponential effect in a
semilog plot, albeit with a factor close to 1.05. 

The results described in what follows were obtained from runs where the basic 
parameters of the model for the infestation had values $M=1000$ and $n=2$. 
The use of different values does not, however, change any of the conclusions, 
which are essentially qualitative.

\begin{figure}[htb]
\centerline{\psfig{file=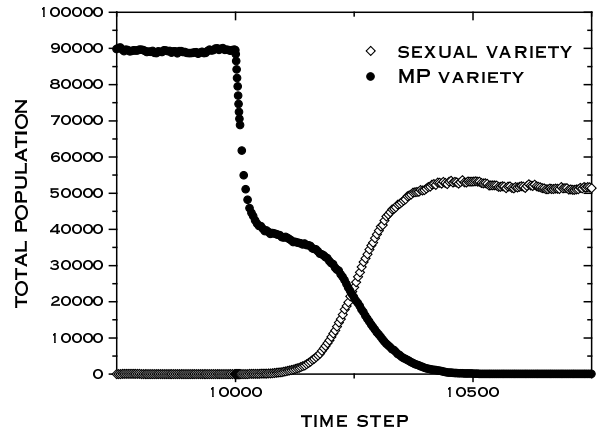, width=9cm, angle=0}}
\caption{Total population of each variety is plotted against time, 
measured in number of Monte Carlo steps. The sexual population is nearly 
vanishing at first, and is kept alive only because of back-mutations from 
the parthenogenetic (MP) one. At time step $10000$ a parasitic infestation 
is turned on, with an exposure of $60$ patterns per female per time step. The 
sexual population rebounds after that, and becomes the dominant variety.}
\label{inf}
\end{figure}

Figure \ref{inf} shows the effect of the parasite infestation. This 
infestation appears at time step $10000$, and is simulated by an exposure 
of each female to $P=60$ genomic imprints of the parasite bank before 
trying to give birth. After having almost vanished, and being kept alive 
solely by the infrequent back-mutations, the sexual variety becomes rapidly 
more abundant as soon as the infestation is unleashed and eventually 
dominates. In accordance with the Red Queen hypothesis, the greatest 
diversity of genomes that sexual reproduction engenders \cite{why} is 
showing one of its measurable consequences.

\begin{figure}[htb]
\centerline{\psfig{file=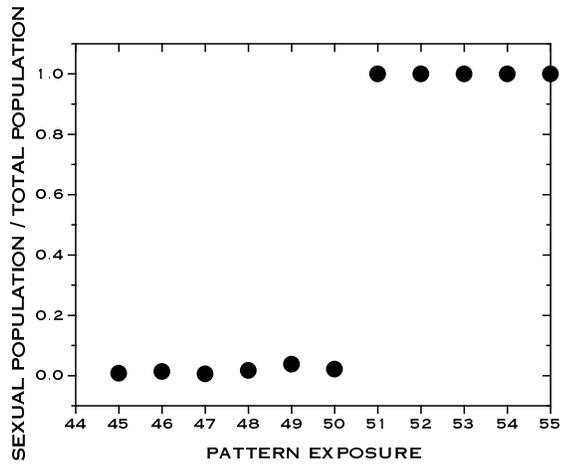, width=9cm, angle=0}}
\caption{Final fraction of sexual population in the total is plotted
against parasite exposure - the number of parasite patterns with which a 
female's genome is compared at each time step. The former plays the role of 
the order parameter of the parasitic-driven transition, while the exposure is
related to parasite density. The sharp jump signals a first-order transition.}
\label{trans}
\end{figure}

The density of the parasite infestation, measured by the number of
exposure patterns, drives a transition between dominating reproductive
regimes. This transition can be measured through the fraction of total
sexual population over the total population. This number acts as an order
parameter for this transition. Figure \ref{trans} shows this fraction as a
function of parasite exposure after a stationary population has been
reached. The sudden jump in the order parameter is signaling a first-order
transition. 
 This claim is further supported by the observation that some runs near
the transition, differing only by the sequence of pseudorandom numbers
used, had unusually long relaxation times. These long relaxations can be
understood as related to metastable states, typical of first-order
transitions.

This paper reports on a simulation of
the coevolution of sexual and parthenogenetic varieties of a same species,
competing for the same resources, in the framework of the Penna bit-string
model. The two-fold disadvantage of having to produce males shows its deadly 
effect on the sexual population, in a static environment. On the other hand, 
the introduction in the model of an infestation of rapidly mutating 
parasites, tailored to reproduce recent observations of existing species, 
can revert the outcome of this competition. This result is in complete 
accordance with those observations, and acts as support for the Red Queen 
hypothesis for sex maintenance in the natural world. A transition, 
conjectured to be of first-order, between dominant reproductive regimes 
observed in nature could be simulated. The selective pressure of a mutating 
ecology is enough to enhance the genetic diversity promoted by sexual 
reproduction, giving it the upper hand against competing asexual populations. 
The results reported show that very simple simulational models can in fact be 
explored as testing ground for theories in biology where observational 
evidence is lacking or insufficient. 

I am greatly indebted to D. Stauffer's quest for eternal youth under the
form of the challenges that he constantly proposes, and to my collaborator
S.M. de Oliveira for drawing my attention to the possibilities of Penna's
model. My work is supported by DOE Grant No. DE-FG03-95ER14499.

\end{document}